\documentclass[twocolumn,showpacs,aps,prl,amsmath,amssymb]{revtex4}

\usepackage{epsfig}

\begin{document}

\title{Pairing mechanism in Fe pnictide superconductors} 
\author{R. Chan${}^{1}$} 
\author{M. Gulacsi${}^{2}$}
\author{A. Ormeci${}^{3}$} 
\author{A. R. Bishop${}^{4}$}
\affiliation{${}^{1}$NICTA, Canberra Research Laboratory, ACT 2612, Australia}
\affiliation{${}^{2}$Dipartimento di Fisica, Universit{\`{a}} di Perugia, Perugia, Italy}
\affiliation{${}^{3}$Max-Planck-Institut f{\"{u}}r Chemische Physik, Dresden, Germany}
\affiliation{${}^{4}$Los Alamos National Laboratory, Los Alamos, NM 87545, USA}

\date{\today}

\begin{abstract}
By applying an \textit{exact} unitary transformation to a two-band
hamiltonian which also includes the effects due to large pnictogen
polarizabilities, we show that an attractive spin-mediated Hubbard term
appears in the $d_{xz}$, $d_{yz}$ nearest-neighbour channel. This pairing
mechanism implies a singlet superconducting order parameter in iron
pnictides.
\end{abstract}
\maketitle

The recent observation of high temperature superconductivity in Fe pnictide
compounds \cite{one} has generated a huge interest and a large research
effort.  As a result four families of compounds have emerged having similar
properties and exhibiting superconductivity.  Namely, chalcogenides (FeSe
and FeTe) and three families of pnictides (represented by LaOFeAs,
BaFe$_2$As$_2$, and LiFeAs).

Band structure calculations have revealed that (i) electron-phonon coupling
alone cannot account for the high values of $T_c$ \cite{two}, (ii) nesting
between the hole Fermi surfaces at zone center and the electron Fermi
surfaces around the zone corner may play a significant role both in
magnetism and superconductivity \cite{three}.  These results have prompted
the idea that the pairing glue is provided by spin fluctuations exchange
between electrons in different bands \cite{three,Kuroki}. However, NMR
studies measuring the temperature dependence of nuclear spin-lattice
relaxation rate $(1/T_1)$ suggest that spin fluctuations may not be
important in iron-pnictide superconductivity \cite{RRR}. Thus other pairing
mechanisms have to be responsible for the unusually high superconducting
critical transition temperature.

In the following we will present the results of an {\sl exact} unitary 
(canonical) transformation (UT) by which it can be shown rigorously 
that an attractive spin-mediated Hubbard term appears in the $d_{xz}$, $d_{yz}$ 
nearest-neighbour channel. The mechanism is similar to a spin double-exchange
type and hence has origins of a kinetic mechanism. 

We start with the by now familiar two-band model description of the
FeAs superconductors \cite{2band}. We take a two-dimensional square lattice
with $d_{xz}$ and $d_{yz}$ orbitals per site.  For simplicity we 
label the $d_{xz}$ and $d_{yz}$ orbitals by $a$ and $b$, respectively.
The kinetic energy component can be expressed as:
\begin{equation}
H_{\rm kin}  = - \frac{1}{2} \sum_{{\bf i}{\bf j}, \sigma}
\sum_{\alpha, \beta = a, b} t_{\alpha \beta}
(\alpha_{{\bf i},\sigma }^{\dagger} \beta_{{\bf j},\sigma }^{} + \textrm{h.c.}) \; .
\label{kin}
\end{equation}
Here, the site indices, {\bf i} and {\bf j}, run over nearest and
next-nearest neighbors. The effective hoppings, $t_{\alpha \beta}$, have
contributions from both direct Fe-Fe and Fe-As-Fe processes.  If the
simplified description of a Slater-Koster formalism \cite{SK-WAH} is adopted,
then $t_{ab}$, the matrix element between $d_{xz}$ and $d_{yz}$ orbitals on
nearest neighbor sites, is zero because the formalism assumes perfect
spherical symmetry. However, in models that use all five Fe 3$d$ orbitals,
$t_{ab}$ is non-zero \cite{Kuroki,ManKax}. In fact, according to Ref. \cite{ManKax}, 
$t_{ab}$ = 0.54 eV is the largest Fe-Fe hopping integral. 
Consequently, we take the view that the two-band model can be
extracted from a more general model by focusing on its relevant $2 \times
2$ block. The parameter $t_{ab}$ is crucial because, as will be discussed
below, it will be at the basis of the UT (see, Eq. 
(\ref{cals})).  The other hopping parameters enter into our UT 
formulation through the bandwidth, $\varepsilon$, which we take as
2 eV \cite{Kuroki,2band,ManKax,hirschfeld}.

$H_{\rm int} = \sum_{{\bf i}} H_{\bf i}$ 
contains only on-site contributions, with
\begin{eqnarray}
H_{\bf i} &= \sum_{\alpha} U_{\alpha} 
n_{{\bf i},\alpha,\uparrow} n_{{\bf i},\alpha,\downarrow} + 
(U' - J/2) \sum_{{\bf i}} n_{{\bf i}, a} n_{{\bf i}, b} \nonumber \\
& - 2 J {\textbf S}_{{\bf i},a} \cdot {\textbf S}_{{\bf i},b} 
+ J' (a^{\dagger}_{{\bf i}, \uparrow} a^{\dagger}_{{\bf i}, \downarrow}
b^{}_{{\bf i}, \downarrow} b^{}_{{\bf i}, \uparrow} + \textrm{h.c.}) \; .
\label{int}
\end{eqnarray}
As in Eq. (\ref{kin}), $\alpha = a, b$ labels $d_{xz}, d_{yz}$ orbitals. 
${\textbf S}_{{\bf i}, \alpha}$ ($n_{{\bf i},\alpha,\sigma}$) is the spin (density) 
in orbital $\alpha$ at site {\bf i}. Following Refs. \cite{2band}, we used 
$U' = U - 2 J$ and the pair hopping term strength $J' = J$, where
$J$ is the Hund coupling. 

The onsite Hubbard terms are obviously equal, $U_a = U_b = U$, with $U$
chosen between 3.0 - 4.5 eV. Although in the literature there is no
agreement on whether the iron pnictides should be considered as strongly
correlated or, at most, moderately correlated materials, there is more
consensus regarding the strength of $U$ to be 4 eV \cite{dmft1},
3.5 eV \cite{weak1} or even smaller $U \ge$ 2 eV \cite{weak2}. 
In our UT approach described below, $U'$ and $J$ give rise to no new
physics, although they do render the transformation algebra very
complicated. Thus, only at the end their effect on the final result
will be briefly discussed. 

The third and last contribution that we include in our starting hamiltonian
is the polarizability effect. 
As pointed out already in Ref. \cite{takahashi} the polarizability effects
in iron pnictides are much larger than in copper-base high $T_c$'s. 
In fact, it is known \cite{fulde} that the ions $As^{3-}$ and $Se^{2-}$ have
huge polarizabilities, mainly due to the fact that their volume is very
large.  Hence, whenever an iron site is charged due to electrons hopping to
or from it, the surrounding As or Se atoms will easily get polarized,
an effect we need to capture.

The effect of As (or Se) polarizability on iron sites can be described
by writing \cite{meinders} the hamiltonian term first introduced in 
Ref.~\cite{deboer} per iron site as 
$g \: n \: (p^{\dagger} s^{} + s^{\dagger} p^{})$. Here the
possible excitations of an As electron from 4$p$ to unoccupied 5$s$ are
taken into account with an effective coupling, denoted by $g$, due to a
charge on Fe.  The $n = n_{\uparrow} + n_{\downarrow}$ notation ($n$ = $n_a$
or $n_b$) is identical to the one used in Eq. (\ref{int}) and describes Fe
3$d$ electrons.

The angle dependence between the As $p$ and $s$ orbitals was extensively studied
already \cite{0811.0214}, hence we only consider a simplified version of 
the hamiltonian taking into account only the mean total polarizability 
per bond \cite{meinders,deboer} by a hamiltonian term of the form:
$H_{\rm pol} = P \sum_{\langle {\bf i},{\bf j} \rangle} (n_{\bf i} - n_{\bf j})$, where
$P = g \langle p^{\dagger} s^{} + s^{\dagger} p^{} \rangle$ will be the
measure of As average polarizability, with $P$ estimated to be
around 2eV \cite{0811.0214}. In this way our starting hamiltonian 
is $H_{\rm kin} + H_{\rm int} + H_{\rm pol}$. 

One of the most common tools used in theoretical physics is perturbation
theory.  Here, we present a perturbation theory which we solve
{\sl{exactly}} using a UT. There are two reasons for using UT \cite{nat_new4}: 
{\sl i}) the belief that the transformed Hamiltonian is 
``simpler'' in the sense that it is ``more diagonal''
and {\sl ii}) the desire to gain a deeper physical 
insight into the problem, given that the transformed 
Hamiltonian may reveal the appropriate independent 
subsystems. Our scope is to pursue the latter. 

For the UT to work, we separate out from $H_{\rm kin}$
the diagonal hopping term, $H_{ab} = t_{ab} 
\sum_{\langle {\bf i}, {\bf j} \rangle, \sigma}
(a^{\dagger}_{{\bf i}, \sigma} b^{} _{{\bf j}, \sigma} + \textrm{h.c.})$. 
Hence we can write 
$H_{\rm kin} = H_{0} + H_{ab}$, where $H_{0}$ contains the remaining terms
other than $t_{ab}$. To be more transparent, we re-write our starting
hamiltonian as $H + H_{ab}$, where 
$H \equiv H_{0} + H_{\rm int} + H_{\rm pol}$. $H$ will be  
the zeroth-order hamiltonian and $t_{ab}$ the perturbation. 

It is known \cite{emery,nat_new6} that in fourth order perturbation an
attractive Hubbard type carrier-carrier interaction appears, parallel to
spin-carrier, spin-spin, carrier-carrier-spin, etc. terms
\cite{fourthorder,hartman}.  All these terms are spin mediated, of
superexchange type \cite{emery,nat_new6}, and hence with physical origins of
a kinetic mechanism \cite{PWA}. The attractive carrier-carrier terms that we
are interested in appear either onsite or between nearest-neighbor sites
\cite{emery,nat_new6}.  However, performing a perturbation up to fourth
order is not enough to draw a final conclusion, as was pointed out in
Ref. \cite{nat_new6}. Depending upon the value of input parameters, the
second, sixth, etc., order terms may well give repulsive interactions, while
the fourth, eighth, etc., attractive ones \cite{hartman}. Thus, higher
contributions need to be calculated in order to verify the convergence.

Hence, initially we check for the convergence of higher order terms
for the particular case of Fe pnictides. In a standard UT  
\cite{nat_new4,emery,nat_new6,fourthorder,hartman} the transformed
Hamiltonian  
$e^{{\cal S}} ( H + H_{ab} ) e^{-{\cal S}}$ 
is identical to 
$H + \sum^{\infty}_{n = 1} [ 1 / n! - 1 / (n+1)!] \: {\widetilde{H}}_{n}$
(henceforth, ${\widetilde{~~}}$ denotes a UT result), where
\begin{equation}
{\widetilde{H}}_{n} \: = \: {\buildrel{n \: {\rm times}}\over
{\overbrace{
[{\cal S}, [{\cal S}, [{\cal S}, \: \cdots \: [{\cal S}}}},
H_{ab}] \: \cdots \: ]]]
\label{ut}
\end{equation}
for an ${\cal S}$ which satisfies $H_{ab} + [{\cal S}, H] = 0$. 
Using the notations:  
${\cal C}_1 = 1/(\varepsilon + P)$, 
${\cal C}_2 = 1/(\varepsilon + P + U_{b}) - {\cal C}_1$,
${\cal C}_3 = 1/(\varepsilon + P - U_{a}) - {\cal C}_1$,
and 
${\cal C}_4 = 1/(\varepsilon + P + U_{b} - U_{a}) 
- 1/(\varepsilon + P + U_{b}) 
- 1/(\varepsilon + P - U_{a}) +  {\cal C}_1$, 
the unitary operator ${\cal S}$ is: 
\begin{eqnarray}
{\cal S} &= - t_{ab} \sum_{\langle {\bf i}, {\bf j} \rangle, \sigma} 
( {\cal C}_1 + {\cal C}_2 n^{b}_{{\bf j}, - \sigma} + 
{\cal C}_3 n^{a}_{{\bf i}, -\sigma} \nonumber \\
&+ {\cal C}_4 n^{a}_{{\bf i}, - \sigma} n^{b}_{{\bf j}, - \sigma}) \:
(a^{\dagger}_{{\bf i}, \sigma} b^{}_{{\bf j}, \sigma} - {\rm h.c.}) \; .
\label{cals}
\end{eqnarray}

\begin{figure}[t]
\includegraphics[width=8.0cm]{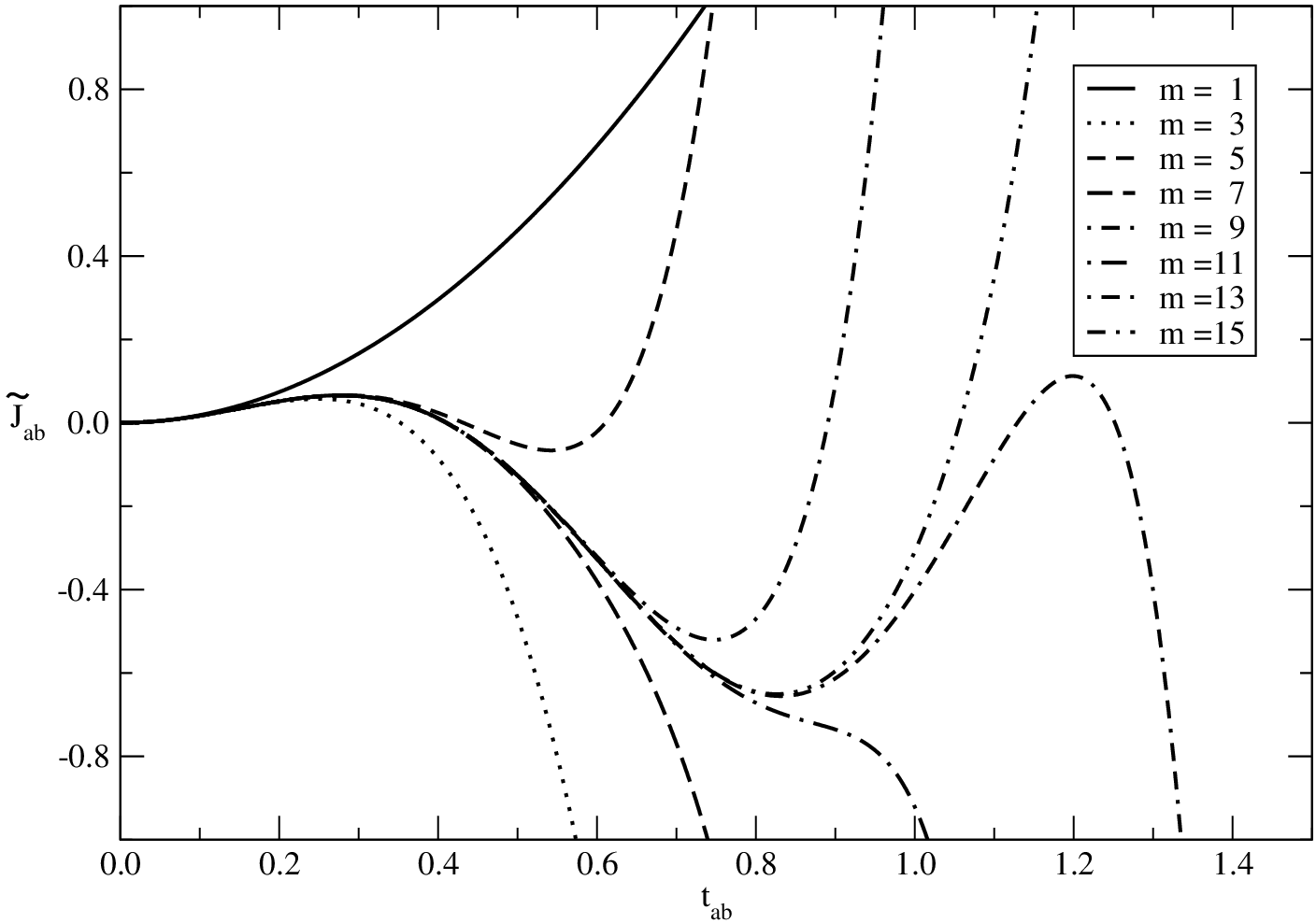} 
\caption{\label{fig:one}
The calculated superexchange interaction ${\widetilde{J}}_{ab}$ in
different orders $n$, as defined in Eq. (\ref{ut}). 
We show the total contributions for each order, i.e., 
$\sum^{\textrm m}_{n = 1} [ 1 / n! - 1 / (n+1)!] \: {\widetilde{H}}_{n}$,
see also Ref. \cite{orders}. For input parameters we used 
$\varepsilon$ = 2eV, $P$ = 2eV and $U$ = 4.5eV. 
}
\end{figure}

The first order UT (i.e., second order perturbation 
theory \cite{nexttoorders}) can be readily performed. This result is well 
known \cite{nat_new6,hartman}, and consequently not pursued further. 
In the second and all other `even' order transformations, $H_{ab}$ 
is recovered structurally with the addition of all possible correlated
hopping terms \cite{nat_new4,nat_new6}: 
$H_{\rm{n = even}} = \sum_{\langle {\bf i}, {\bf j} \rangle, \sigma}  
[ {\widetilde{t}}^{}_{ab}  + {\widetilde{t}}^{a}_{ab} n^{a}_{{\bf i}, -\sigma} 
+ {\widetilde{t}}^{b}_{ab} n^{b}_{{\bf j}, - \sigma}  
+ {\widetilde{t}}^{ab}_{ab} n^{a}_{{\bf i}, - \sigma} 
n^{b}_{{\bf j}, - \sigma}] \: 
(a^{\dagger}_{{\bf i}, \sigma} b^{}_{{\bf j}, \sigma} + {\rm h.c.})$. 

Consequently, the weight of the original $t_{ab}$ is redistributed 
in every order of the transformation among 
${\widetilde{t}}^{}_{ab}$, ${\widetilde{t}}^{a}_{ab}$, 
${\widetilde{t}}^{b}_{ab}$ and ${\widetilde{t}}^{ab}_{ab}$. 
Because of the smallness of $U$ in Fe pnictides, 
as the UT is performed to higher orders weight
is shifted in and out of the correlated hopping terms
generating an oscillatory behavior. These oscillations
are well-known \cite{nat_new4,nat_new6,hartman} in standard perturbation
theory. As an example, the Heisenberg superexchange term ${\widetilde{J}}_{ab}$ 
is shown in Fig. \ref{fig:one} for high orders. 
Hence, the problem with applying standard UT 
is that unless $U$ is very large, the perturbation series will not converge. 

To overcome this problem, a completely new approach is needed to 
handle Fe pnictide case, i.e., to perform an exact UT.
We have chosen to perform such a transformation by eliminating three
consecutive even order terms simultaneously 
(the $n=0$, $n=2$ and $n=4$ order terms) from Eq. (\ref{ut}) via a new
unitary operator ${\cal S}$ such that 
\begin{equation}
H_{ab} + [{\cal S},H] + 
\frac{1}{2!} [{\cal S},[{\cal S},H_{ab}]] \: = \: 0 \; . 
\label{uj_ketto}
\end{equation}
This guarantees that the transformation cannot be continued to higher
orders, since all $a \leftrightarrow b$ hopping processes are 
eliminated, as ${\widetilde{t}}^{}_{ab}$ and the correlated
hopping terms, ${\widetilde{t}}^{a}_{ab}$, ${\widetilde{t}}^{b}_{ab}$, 
${\widetilde{t}}^{ab}_{ab}$ are strictly zero.  Accordingly, the transformed
hamiltonian 
\begin{equation}
e^{{\cal S}} ( H + H_{ab} ) e^{-{\cal S}} \: = \:
H + [{\cal S},H_{ab}] + 
\frac{1}{2!}[{\cal S},[{\cal S},H] \; ,
\label{uj_egy}
\end{equation}
is exact in a strict mathematical sense \cite{nat_new4}. 
Eq. (\ref{uj_egy}) is merely a self-consistent mixing \cite{nat_new4} of the n=1 
and n=3 order standard UT, i.e., of a standard 
2nd and 4th order perturbation theory \cite{nexttoorders}. 
This guarantees that we can capture the whole spectrum of the carrier-carrier 
interaction terms \cite{emery,nat_new6}. 

\begin{figure}[t]
\includegraphics[width=8.0cm]{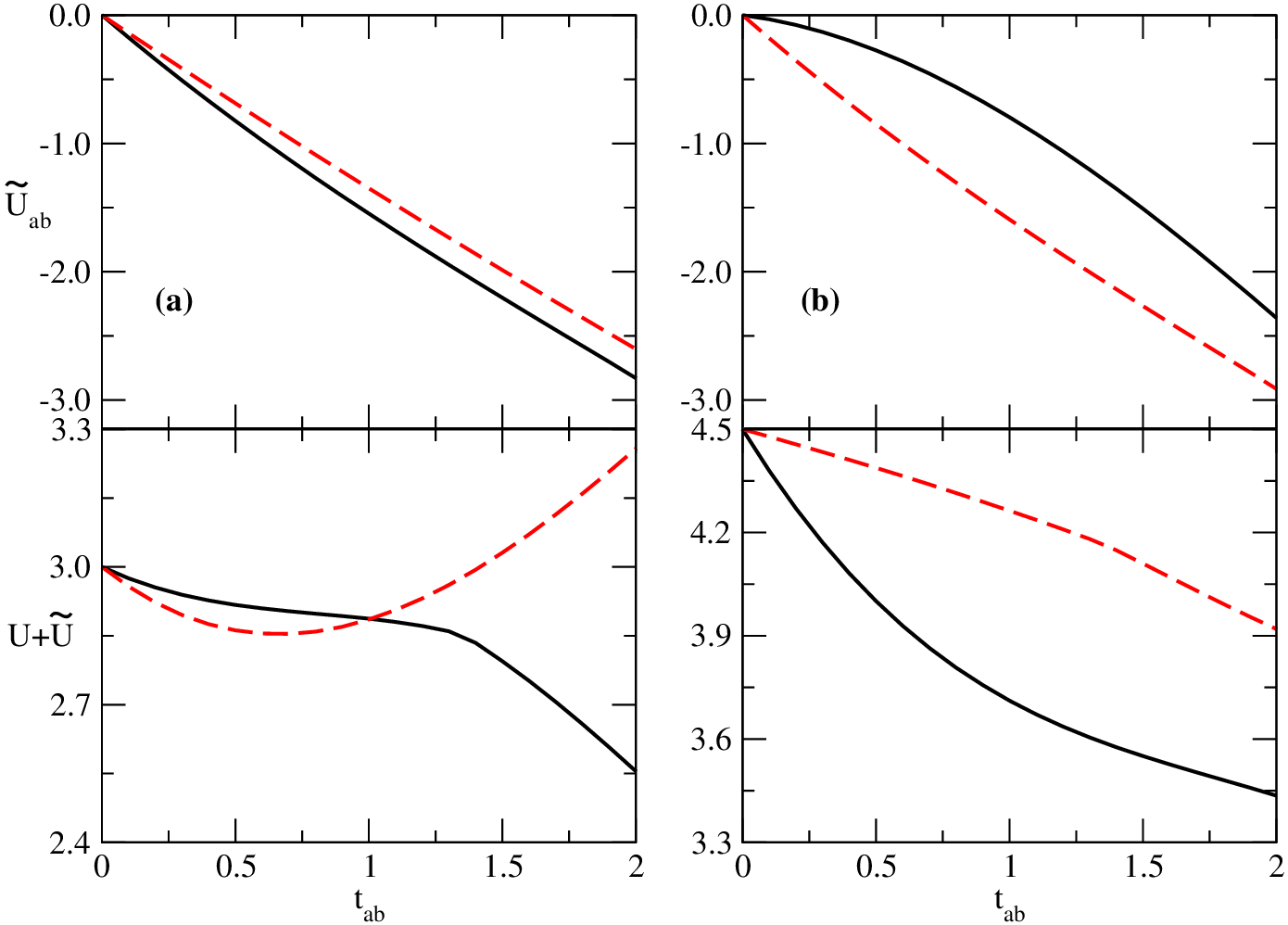} 
\caption{\label{fig:two}
(Color online) The calculated nearest neighbor, ${\widetilde{U}}_{ab}$ (top), 
and the total onsite, $U$ (initial value) + ${\widetilde{U}}$ (calculated value)
(bottom), Hubbard coefficients as a function of $t_{ab}$. Input parameters
are $\varepsilon$ = 2eV, $U$ = 3eV [part (a)] and $U$ = 4.5eV [part (b)]. 
Continuous (black) curves are $P$ = 1.5eV, while long-dashed (red) curves 
are for $P$ = 3eV. 
}
\end{figure}

The solution of Eq. (\ref{uj_ketto}) has the same algebraic form as  
Eq. (\ref{cals}) with new coefficients ${\cal C}_1$, ${\cal C}_2$, 
${\cal C}_3$ and ${\cal C}_4$ as unknown parameters. These are
determined from Eq. (\ref{uj_ketto}), which gives the following 
systems of equations:

${\widetilde{t}}^{}_{ab} = t_{ab} [ \cos(2 t_{ab} {\cal C}_1) +  
(\varepsilon + P) {\cal C}_1 
\sin(2 t_{ab} {\cal C}_1) / (2 t_{ab} {\cal C}_1) ]$

${\widetilde{t}}^{a}_{ab} = 8 t_{ab}^{3} \mu 
[ (\varepsilon + P) (\mu^{2} - \nu^{2}) - U_a \mu^{2} - U_b \nu^{2} ]
\phi(2 \tau) 
+  8 t_{ab}^{3} \mu (\mu - \nu) \psi(2 \tau) 
+  4 t_{ab}^{3} \mu \nu^2 (\varepsilon + P) \phi(\tau) 
+  2 t_{ab}^{3} \nu (\mu + \nu)\psi(\tau) 
-  t_{ab} \cos ( 2 t_{ab} {\cal C}_1 )
- (1/2) (\varepsilon + P) \sin ( 2 t_{ab} {\cal C}_1 )$, 
where $\mu = {\cal C}_1 + {\cal C}_2$, $\nu = {\cal C}_1 + {\cal C}_3$,
$\tau = t_{ab} [2(\mu^2 + \nu^2)]^{1/2}$ and
the notations $\phi (x) = \sin (x) / x^3$, $\psi (x) = \cos (x) / x^3 $
were used. For ${\widetilde{t}}^{b}_{ab}$ we have to exchange $U_a$ with $U_b$. 
Finally, the last of the correlated hopping terms is: 

${\widetilde{t}}^{ab}_{ab} = - 8 t_{ab}^{3} (\mu - \nu)  
[ (\varepsilon + P) (\mu^{2} - \nu^{2}) - U_a \mu^{2} - U_b \nu^{2} ]
\phi(2 \tau) -  8 t_{ab}^{3} (\mu - \nu)^2 \psi(2 \tau) 
-  2 t_{ab}^{3} \mu \nu (\mu + \nu) 
[(\varepsilon + P) - U_a + U_b] \phi(\tau) 
-  2 t_{ab}^{3} (\mu + \nu)^2 \psi(\tau) 
+ t_{ab} ({\cal C}_1^2 + \rho^2) \cos (2 t_{ab} \rho) 
+ \frac{1}{2} (\varepsilon + P) (1 + {\cal C}_1^3 / \rho^3) 
\sin ( 2 t_{ab} \rho )
+ t_{ab} (1 - {\cal C}_1^2 / \rho^2) 
[ 1 + (\varepsilon + P) {\cal C}_1 ]$,
where $\rho = {\cal C}_1 + {\cal C}_2 + {\cal C}_3 + {\cal C}_4$ and note that
these equations can also be used if $U_a \ne U_b$. 

\begin{figure}[t]
\includegraphics[width=8.0cm]{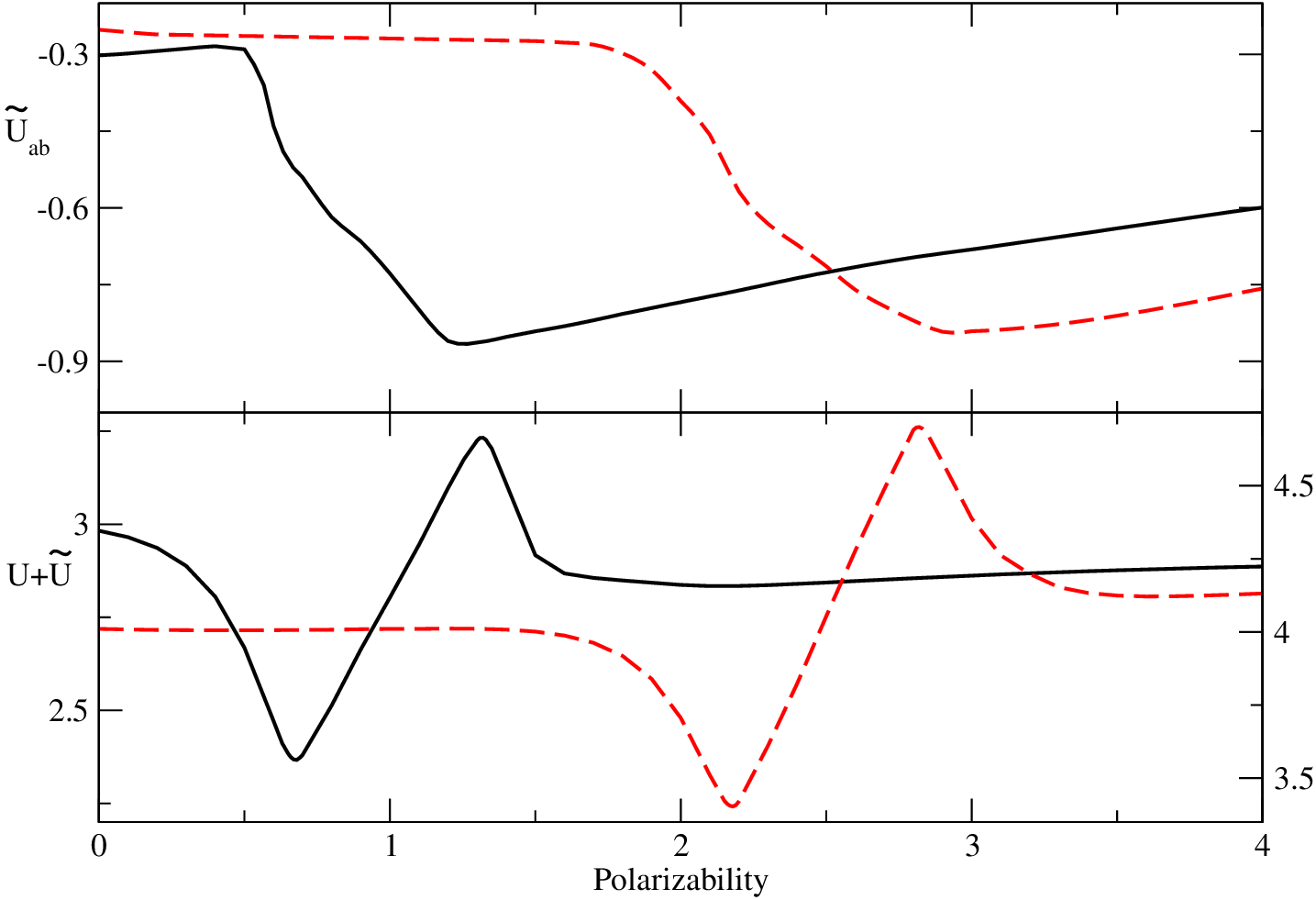} 
\caption{\label{fig:three}
(Color online) The same Hubbard coefficients as in Fig. \ref{fig:two} as a 
function of the polarizability, $P$, for $\varepsilon$ = 2eV and $t_{ab}$ = 0.5eV.
$U$ = 3eV, continuous (black) curves and 4.5eV long dashed (red) curves, respectively. 
For the lower curves, the left (right) axis is for $U$ = 3eV (4.5eV).   
}
\end{figure}

We calculated numerically the parameters 
${\cal C}_1$, ${\cal C}_2$, ${\cal C}_3$ and ${\cal C}_4$ 
from the above system of equations by setting ${\widetilde{t}}^{}_{ab}$, 
${\widetilde{t}}^{a}_{ab}$, ${\widetilde{t}}^{b}_{ab}$ and 
${\widetilde{t}}^{ab}_{ab}$ to zero. 
The new UT, as any standard perturbation theory, gives 
several spin-spin, carrier-carrier, spin-carrier onsite, nearest neighbor,
next-nearest 
neighbor, etc. terms. From all these we only calculated explicitly the onsite, 
${\widetilde{U}} \sum_{{\bf i}} n_{{\bf i}, \alpha, \uparrow} 
n_{{\bf i}, \alpha, \downarrow}$ ($\alpha = a, b$) and nearest neighbor 
${\widetilde{U}}_{ab} \sum_{ \langle {\bf i}, {\bf j} \rangle} 
n_{{\bf i}, a, \uparrow} n_{{\bf j}, b, \downarrow}$
Hubbard interactions. The results are plotted in Fig. \ref{fig:two}, as a
function  of $t_{ab}$, and in Fig. \ref{fig:three} as a function of the
polarizability, $P$.

As can be seen in Fig. \ref{fig:two}, a nearest-neighbor inter-orbital
attractive Hubbard interaction ${\widetilde{U}}_{ab}$ appears due 
to a spin-mediated superexchange type mechanism \cite{emery,nat_new6}.
Being of kinetic origin, the spins are not excited \cite{PWA}, only virtual
excitations of onsite singlets occur. The situation is similar
to the phonon-mediated attraction: at temperatures well below the Debye
scale real phonons are never excited, yet they provide an attraction
mechanism between electrons.

The values of ${\widetilde{U}}_{ab}$ in Fig. \ref{fig:two} can be well 
approximated for small $t_{ab}$ with a line: 
${\widetilde{U}}_{ab} \approx - \; {\rm{const}} \; t_{ab}$, 
with the constant being $4 \xi (\zeta_{+} - \zeta_{-})$ 
$\{ \sin (2 [2 \xi^2 (\zeta^2_{+} + \zeta^2_{-})]^{1/2}) / 
(2 [2 \xi^2 (\zeta^2_{+} + \zeta^2_{-})]^{1/2})$ + 
$[ \cos(2 [2 \xi^2 (\zeta^2_{+} + \zeta^2_{-})]^{1/2}) - 1 ] / 
(2 [2 \xi^2 (\zeta^2_{+} + \zeta^2_{-})]^{1/2})^2 \}$, where
$\zeta_{\pm} = 1/[\varepsilon + {\textrm P} \pm U]$,  
$\xi = \eta \vert (\varepsilon + {\textrm P})^2 - U^2 \vert /
[ (\varepsilon + {\textrm P})^2 + U^2 ]^{1/2}$ and the numerical
coefficient $\eta$ is 0.392 for $\varepsilon + {\textrm P} < U$,
or 1.178 otherwise. 

The new hamiltonian term, $H_{\rm Pol}$, which gives a measure of the 
polarizability is not crucial in obtaining this attractive interaction.  
Any finite value of bandwidth alone suffices for attraction. In fact, the UT
expressions depend only on the sum $\varepsilon + P$, and not separately on
either. Thus, polarizability effectively acts as a \textit{bandwidth}. So,
Fig. \ref{fig:three} covers a wider range for $P$ to allow for cases with
different bandwidth and $P$ values. From Fig. \ref{fig:three}, upper panel
we observe that the attractive ${\widetilde{U}}_{ab}$ is enhanced rather
strongly by $P$ for $\varepsilon + P \gtrapprox U$. The strongest attraction
occurs around the values of $P$ satisfying $\varepsilon + P \approx U$.
Note that, this regime is not accessible by standard perturbation theory, but
it works well in an exact UT approach. 

An additional effect due to $H_{\rm Pol}$ is the modified screening of the
on-site Coulomb repulsion $U$ in the presence of $P$. In the lower panel of
Fig. \ref{fig:three} the net repulsion, $U + {\widetilde{U}}$, is seen to
deviate significantly from a roughly constant value ($\le 3$ eV for $U=3$
eV, $\approx 4$ eV for $U=4.5$ eV) only in an interval given by
$| \varepsilon + P - U | \le 0.5$ eV $= t_{ab}$. Normally, ligand
polarizabilities are expected to increase the screening (smaller net $U$
values), however, when the double exchange is the strongest, it restricts
the hopping and as such screening is reduced, similarly to, e.g., manganites
\cite{fulde}. $U'$ and $J$ from Eq. (\ref{int}), have been neglected up to now. 
However, their effects in our UT can be accounted for by substituting
everywhere $U$ with $U + J ( 1 - U' / U)/2$. Using the known $U'$
and $J$ values \cite{Kuroki,2band,ManKax,hirschfeld,dmft1,weak1,weak2} this only
causes a less than 5\% change in $U$ and hence their effect is minor.

In conclusion we have shown that the Fe pnictide family of alloys exhibit
a rather strong pairing interaction, due to a spin-mediated, superexchange type
mechanism. There is a quite strong enhancement of 
the attraction due to the polarizability. However we do not observe polaron and
bipolaron formation as in Ref. \cite{0811.0214}, probably because we are using 
realistic onsite Hubbard values, instead of $U > 8$ eV as used in Ref. \cite{0811.0214}. 
Our pairing mechanism will give rise to a singlet superconducting order parameter. 
However, in a two-band system the self-consistent solution of the gap equations 
always has a symmetric (recently denoted as $s_{++}$) and asymmetric 
($s_{+-}$) solution \cite{rusianov}. In the parameter regime of superconducting 
Cr alloys, at different concentration of electrons and holes, the asymmetric 
solution wins \cite{us}. In closing we remark that non-phononic 
mechanisms support \cite{penn} negative isotope effect \cite{0903.3515}. 
The standard isotope effect measured \cite{isotope_effect} for both
superconductivity and  itinerant antiferromagnetism in Fe pnictides would
require phonon input, which may contradict an electronic mechanism. However,
this may not necessarily be the case, as phonons can act as random impurity
potentials in certain cases for systems exhibiting electron and hole Fermi
surfaces \cite{barker}. The pair-breaking effect originating 
from this also explains why Cr has an isotope effect in $T_N$. 
This actually can be very large; the phonons can decrease $T_N$ 
by as much as 70\% from its value in their absence \cite{cr_isotope}. 
Another interesting consequence of the pair-breaking
by virtual or thermal phonons \cite{cr_isotope} is that the ratio
$2 \Delta / k_B T_c$ can be much larger than the BCS value of 3.53. 

MG thanks L. Hozoi and P. Fulde for fruitful discussions. He also  
acknowledges the financial support of the Max Planck Institute for the Physics 
of Complex Systems, Dresden, where most of this work was carried out.


\begin{thebibliography}{99}
\bibitem{orders}As the number of terms ($2^{1 + n + (n-3)/2} 5^{(n-1)/2}$,  
$n \ge 3$) which contribute to ${\widetilde{J}}_{ab}$ 
in each individual odd order can reach tens of billions, we used a purpose
built algebraic manipulation program to calculate higher orders.
\end{thebibliography}
\end{document}